**Ana Vukadin,** National and University Library Zagreb, Croatia
**Aida Slavic,** UDC Consortium, The Netherlands

# Challenges of Facet analysis and Concept Placement in Universal Classifications: the Example of Architecture in UDC

**Abstract**
The paper discusses the challenges of faceted vocabulary organization in universal classifications which treat the universe of knowledge as a coherent whole and in which the concepts and subjects in different disciplines are shared, related and combined. The authors illustrate the challenges of the facet analytical approach using, as an example, the revision of class 72 in UDC. The paper reports on the research undertaken in 2013 as preparation for the revision. This consisted of analysis of concept organization in the UDC schedules in comparison with the Art & Architecture Thesaurus and class W of the Bliss Bibliographic Classification. The paper illustrates how such research can contribute to a better understanding of the field and may lead to improvements in the facet structure of this segment of the UDC vocabulary.

## 1. Introduction

The research reported in this paper is occasioned by the revision of class 72 *Architecture* in the Universal Decimal Classification (UDC) that started in spring 2013. UDC is used in many specialised collections in the domains of art, architecture and the building industry. Numerous in-house schemes in specialised libraries of this type were derived from the UDC and some compilations of UDC subjects were published as books (e.g. ABC - Abridged Building Classification for architects, builders, civil engineers 1981).

The ongoing revision of class 72 is initiated to meet the demands of these collections by re-structuring, consolidating, updating and expanding the vocabulary in architecture and all related subjects: physical planning, civil engineering and building industry. In doing so it was envisaged that UDC may also become more aligned with other widely used vocabulary standards for these subjects, in particular Art and Architecture Thesaurus (AAT). In this paper we report only on a small part of the initial research which may be of interest to the wider knowledge organization community.

## 2. Facet analytical theory and UDC

The UDC scheme standard known as UDC Master Reference File (UDC MRF) was created in 1992, based on a selection of 60,000 out of 200,000 subdivisions of the previous full UDC edition. As a consequence many subject areas have lost more specific terminology. The then UDC editorial plan was to continue with the development of the schedules based on a more manageable UDC MRF by implementing a rigorous and consistent analytico-synthetic principle. This meant that facets of concepts would be developed in such a way that they would be shared between different areas of knowledge with no need for repetition or enumeration of compound concepts. Thus the scheme would become more powerful and flexible in indexing while the growth of the vocabulary remains controlled. In subsequent years many subject areas were revised following this principle (Slavic, Cordeiro & Riesthuis 2008).

UDC falls into a category of classifications termed by Ranganathan (1965, 86) as 'guided analytico-synthetic schemes'. The principle of synthesis in UDC allows for the combination of all subjects between themselves or the extension of all subjects with facets of *common auxiliaries* (place, time, persons, materials etc.) or the combination with *special auxiliaries* (most typically facets of operations, agents, tools etc.) that can be shared between groups of subjects. UDC's important feature is that, although there is a provision for guided ordering and pre-coordination of concepts, the use of the scheme does not depend on a strict order of facets. The notational system in UDC allows easy building and parsing of complex notations which offers great flexibility in the indexing and arrangement of subjects. In principle, the UDC vocabulary structure supports coordinated indexing and searching more akin to a freely faceted classification (Gardin 1965).

The facet analytical theory and faceted classifications, especially those drawing from the work on the Classification Research Group (CRG), are more often concerned with special subject classifications (e.g. Vickery1960, Broughton & Slavic 2007). In such cases, facet analytical theory is used as a method for a single domain analysis and yields a vocabulary which is not influenced by the treatment of the same concepts in other areas of knowledge. Universal classification schemes, when they function as a coherent knowledge system, such as UDC, represent a challenge when it comes to the application of the CRG-type of facet analysis. This is due to their need to manage the placement, association and repetition of concepts across the universe of knowledge. The UDC revision policy is, therefore, concerned not only with domain facet analysis, but also with determining a unique place and notational representation for a concept, avoiding concept enumeration, repetition or parallel divisions.

The implementation of facet analysis in UDC requires additional effort when it comes to coordinated re-use of concepts and their notational representation across different domains. In addition, each subject represents its own challenge and requires research into methods and approaches which is reflected in the abundance of literature on this topic (Broughton 2000, Williamson & McIlwaine 2009, Gnoli 2009). The arts and humanities were often recognized as an especially challenging field for the classificationists (Ørom 2003, Vukadin 2006, Broughton & Slavic 2007). Although theoretical issues in constructing faceted schemes are beyond the scope of this paper we believe that our experience with the revision of architecture will shed some light on the complex issues of the structuring of a faceted scheme.

### 3. Architecture in UDC

The field of architecture is a good example of a pervasive subject which is connected to many other subjects and concepts in a universe of knowledge. This field can borrow, share or provide terminology for many other subject areas. For example, architecture may:

- require facets of general and context-free concepts such as place, time, materials, persons, ethnic grouping, properties, processes etc. that are common to all fields of knowledge;
- apply methods, techniques, tools from other fields of knowledge (e.g. computer science, mathematics, earth sciences, industry);
- share facets of concepts with the arts, landscaping, urban planning, interior design, civil engineering, and the building industry;
- provide basic terminology, such as types of buildings, that is required in many subject fields (e.g. public building, shops, schools, etc.);
- be a subject of study in many areas of knowledge such as social sciences, humanities or technology.

The complexity of the subject area of architecture is reflected in the distribution of architectural concepts within class 7 and across the main classes in UDC (Figure 1).

Figure 1: Organization of architecture and related subjects in UDC

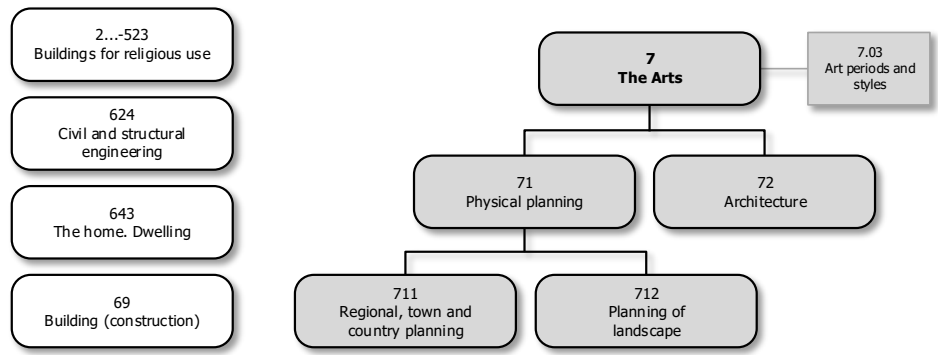

In general, UDC is well suited for indexing and retrieval of architectural subjects: all fundamental domain concepts are present in the scheme and their hierarchical relationships are well represented and supported by synthetic principle which allows very detailed subject description and less dependence on a rigid hierarchical structure. However, there are some important issues precipitating the revision of this subject.

Certain facets of notable relevance for the architectural domain, such as design (movable, temporary buildings, etc.) or environmental conditions (underground buildings, hillside buildings, etc.) are found exclusively as special auxiliaries in class 69 *Building (construction) trade. Building materials. Building practice and procedure*. Built complexes such as memorial sites or recreational areas, which equally belong to the fields of architecture and environmental planning, are divided between classes 71 *Physical planning. Regional, town and country planning. Landscapes, parks, gardens* and 72 *Architecture*.

While this approach technically works, given the analytico-synthetic nature of UDC, the scattering of concepts creates difficulties for indexers, impedes predictability in indexing and searching and may affect the logical arrangement of subjects. Clearly, this subject in UDC would benefit from mechanisms which would allow for a logical, predictable and more direct synthesis of the domain-related concepts.

Another well known issue in this class is a discrepancy between the more recent theory of architecture and the UDC structure which is based on theories and trends in architecture from the first half of the 20th century. This can be observed, for instance, in 711.4 *Urban planning* which is based on the functionalistic view of architecture (Vukadin 2006, 76). Given the synthetic nature of UDC, we need to examine whether this at all affects the functioning of the scheme in indexing and if so how this problem may be addressed.

Finally, the 1992 reduction of the UDC schedule and subsequent revision in 2003, have affected class 72 *Architecture* significantly (Extensions and Corrections to the UDC 2003, 111-119). Over 80% of more specific concepts dealing with architectural elements or buildings and structures were removed or were scattered in other areas of knowledge in the process of subsequent 'cleaning' of concept enumeration and duplication[1]. There is no doubt, therefore, that the expansion of the vocabulary is one of the most important reasons for the revision of this class.

**4. Approach to the revision of architecture**

The starting point of this research was the analysis of the art, architecture and related subject fields in the old full English edition of UDC (BS 1000 [624]:1981; BS 1000 [69]:1981; BS 1000 [72]:1975. The full schedules were designed by an international team of subject specialists for use in special collections and contain very detailed subdivisions. These editions were then compared with the UDC MRF11 to assess the subsequent loss of terminology.

In order to establish the premises for an improvement, we then compared methods of facet organization in architecture in UDC and two other knowledge organization schemes, BC2 (Mills & Ball 2006, Class W) and AAT (Art & Architecture Thesaurus Online 2013). All three systems are generally described as faceted knowledge organization systems: UDC and BC2 are universal classification schemes while AAT is an alphabetical indexing language, thus primarily concerned with linguistic aspects of vocabulary control.

**5. Comparison of UDC, AAT and BC2**

From the outset it was clear that the difference between UDC and the other two systems is significant when it comes to the arrangement of subjects and the scope or coverage of the vocabulary. UDC separates pure sciences and applied sciences while BC2 alternates them and this approach to the relationships between theory and application is also reflected in the arts. In addition both AAT and BC2 appear to deal with architecture as a special subject, i.e. they are not concerned with repetition or re-use of the same concepts in other fields of knowledge (e.g. engineering or industry) or the relationship between concepts across disciplines.

The following basic facets are common to all three systems: types of buildings, spatial and constructive parts of buildings, materials and techniques, processes, and styles or periods. The comparison, however, helps in identifying different approaches in facet organization as is shown in Figures 2 and 3 with a rough mapping of relevant areas between AAT, BC2 and UDC. AAT and BC2 subjects are shown as text - and corresponding UDC classes indicated by notation below.

As a discipline-oriented scheme, UDC is characterized by a stronger distinction between related fields such as environmental planning, architecture and civil engineering, while the other two systems, especially AAT, provide a more holistic view of environmental design (including architecture), permitting the facet organization to be based on inherent criteria such as spatial relationships or functions instead of disciplines. The focus on the relationships between built objects and the environment also resonates with more recent tendencies in architectural theory (e.g. the phenomenology of architecture[2]).

---

[1] Comparison of the number of subdivisions of the full UDC edition from the 1980s and UDC MRF 2012 illustrates this problem very well. For instance, class 72 *Architecture* now has 115 subdivisions (full ed. had 900); class 624 *Civil engineering* has 551 subdivisions (full ed. had 1,154), class 69 *Building* has 705 subdivisions (full ed. had 1,580).

[2] In the 1980s, phenomenological thought had a great influence on architectural schools, primarily through the introduction of the concepts 'genius loci' and 'cultural landscape' (Norberg-Schulz 1979).

The main feature of facet organization in UDC revolves around the facet of 'building and structures according to function' (civic buildings, residential buildings, etc.). Other facets, represented by special auxiliaries (some of which are shared by the whole class 7 *Arts. Recreation. Sport*), are:
- theory and philosophy of architecture (including design and composition)
- techniques
- styles and periods
- architectural details and finishes
- parts and spaces for specific uses (entrances, ancillary spaces, etc.).

Materials and constructive features, as well as additional design features, can be expressed using combinations with concepts 'borrowed' from other classes, notably environmental planning and civil engineering.

Figure 2: Concept organization in AAT in relation to UDC

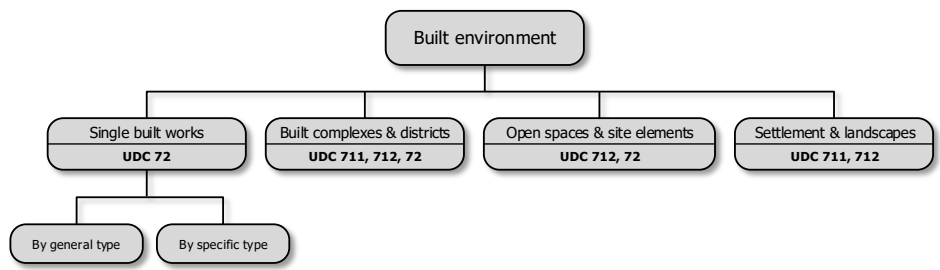

In AAT the hierarchy caption *Built environment*, which is part of the *Objects* category, serves as a unifying heading for four main facets: *Settlements and landscapes, Built complexes and districts, Single built works*, and *Open spaces and site elements*. The facet closest to the main UDC facet of 'building types', *Single built works*, is further divided by general type into buildings (intended to be used for sheltering an activity or occupancy) and structures (which may or may not provide enclosure or shelter, e.g. signal towers). Single built works are also divided by a number of specific type sub-facets:
- by form (height, shape);
- by function (residential, religious, agricultural, military, etc.);
- by condition (abandoned, illegal, etc.);
- by location or context (including topographical location: artificial islands, marine structures, etc.);
- by design (energy-efficient buildings, portable structures, floating structures, etc.);
- by ownership (animal architecture).

Other relevant concepts can be found in respective facets such as *Components* (e.g. parts of buildings), *Conditions and effects* (e.g. corrosion) or *Furnishing and equipment*.

Similarly, BC2 encompasses urban planning, landscaping, civil engineering and architecture under the umbrella caption *Environmental and landscape design*, which belongs to the subclass of *Design arts*. On the other hand, compared with AAT, the general facet organization of BC2 regarding architecture is more aligned with UDC due to its reliance on fields of knowledge, not on objects. The following general facets have been identified:
- composition

- styles and movements in architecture (by period, by place)
- materials of buildings
- parts of buildings
- kinds of buildings

Figure 3: Concept organization in BC2 in relation to UDC

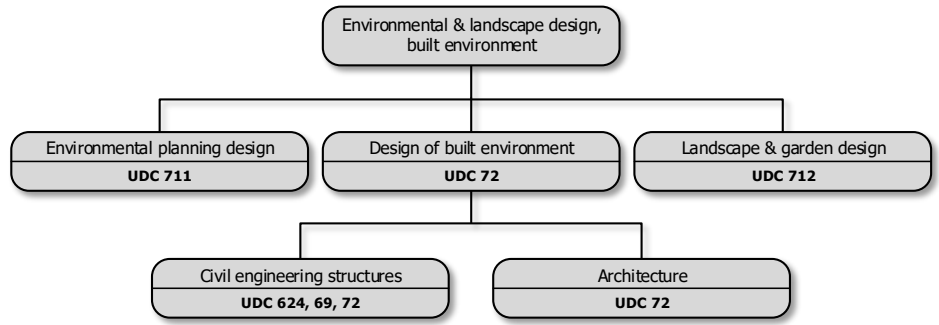

Similarly to AAT, kinds of buildings are sorted according to more than one sub-facet:
- by construction process (e.g. modular system, prefabricated, etc.)
- by material
- by form (e.g. single-story buildings, tall buildings, etc.)
- by attachments (e.g. detached, semi-detached, etc.)
- by support characteristics
- by service-life (e.g. temporary buildings)
- by special purpose (e.g. automated, sustainable buildings, etc.)
- by function (e.g. commercial, residential, administrative buildings, etc.)

A closer inspection of AAT and BC2 shows that they share with UDC all the basic facets of the architectural domain such as form, styles/periods, parts of buildings, or techniques/activities. The most significant differences observed are: a) AAT and BC2 provide more detailed subdivisions of building types according to various criteria; and b) in these two systems the vocabularies of architecture, environmental planning, landscape and design are better integrated than is the case with UDC. These are precisely the areas of UDC in which we have identified the greatest need and scope for improvement.

**6. Impact on UDC revision**

The main decision in designing an analytico-synthetic scheme is to decide what is the main facet around which all other facets of a vocabulary revolve. In spite of the fact that both AAT and BC2 have different approaches, it seems that there is no strong reason to change the nature of the main facet in UDC which is *buildings and structures according to function*. The notion of functionality (or programme) of a building, central to modernist thought, has been questioned by many architects and theorists from the 1960s onward. However this approach has proven to be useful in organizing the domain and most importantly it corresponds to the still prevailing way of representing architectural projects in specialised literature (Vukadin 2006). As it transpires from our research this facet in UDC also ought to be better structured and expanded with more detail. However, instead of introducing sub-facets of *buildings according to design*, *form*, *environmental context*, etc. (as in AAT and BC2), the vocabulary for further specification in UDC can be provided through the existing common auxiliaries of place, materials and, especially, general properties (shape, structure, arrangement etc.).

In the course of this research, it has also become clear that the UDC scheme would benefit if all building types (currently dispersed in other subject areas e.g. agricultural building) would be collocated in class 72 *Architecture*. The unique definition of concepts and unique notational representation allow for more predictable arrangement and retrieval of building types across the scheme. The UDC's analytico-synthetic feature enables any other subject to utilise concepts listed under architecture. This approach is also in line with the growing tendency in architectural practice, as well as in the heritage conservation, to blur the boundaries between what was once considered 'architecture' and what were seen as functional buildings (e.g. barns, windmills, industrial complexes, etc.)

In order to consolidate the shared vocabulary between civil engineering, building industry and home economics it is possible to introduce, under 72 *Architecture,* all shared facets of concepts such as spatial, constructive and functional parts of buildings. This would allow for a stronger synthesis of related domains, while preserving the disciplinary context and reducing the pressure on overcrowded class 6.

Finally, our research into AAT and BC2 influenced our decision to take steps towards the integration of physical planning and architecture by establishing a unifying class above UDC classes 71 and 72 which can accommodate facets of shared concepts presented as special auxiliaries. This would help in expressing a more contemporary view on the built environment, while also providing the possibility of specifying the disciplinary context.

**7. Concluding remarks**

The revision of a subject domain in a widely used and well established scheme such as UDC is a long and complex process, consisting of many phases and several 'back-to-the-drawing-board' steps interjected with editorial reviews and discussion.

The starting assumption of our analysis was that the UDC's facet structure will differ significantly from the AAT and BC2 class W which are both created as vocabularies operating within a single field: the arts. Our objective was to see whether what we learn by observing these vocabularies may help us in improving and expanding UDC.

What we observed both about the similarities and the differences between the systems was very valuable for the preparatory phase of the revision. From our experience it is clear that the analysis of the idea plane of other systems can help to identify both strengths and weaknesses of one's own system and help in providing arguments for new solutions.